\documentclass[11pt, preprint]{aastex}    % Preprint.

\def\kms{$\,$km$\,$s$^{-1}$}
\def\ms{$\,$m$\,$s$^{-1}$}
\def\sqig{$\sim$}

\def\degrees{$^{\circ}$}
\def\jbis{Journal of the British Interplanetary Society}

\shortauthors{R.H.D. Corbet}

\begin{document}

\title{
Synchronized SETI - The Case for ``Opposition''}

%\title{\today}

\author{Robin H. D. Corbet}
\affil{Universities Space Research Association\\
7501 Forbes Blvd, Suite 206\\
Seabrook, MD 20706}
\email{corbet@gvsp.usra.edu}

\begin{abstract}
If the signals that are being sought in SETI programs do exist but are
very brief, for example because they are produced intermittently
to conserve energy, then it is essential to know when these signals will
arrive at the Earth.  Different types of synchronization schemes are
possible which vary in the relative amount of effort which is required by
the transmitter and the receiver. Here the case is made for a scheme which
is extremely simple for the receiver: make observations of a
target when it is at maximum angular distance from the Sun
(i.e. ``opposition'').  This strategy requires the transmitter to
have accurate knowledge of the distance and proper motion of the Sun
and the orbit of the Earth.  It is anticipated that within about the next
10 to 20 years it will be possible to directly detect nearby extra-solar
planets of approximately terrestrial mass. As any extraterrestrial
transmitters are expected to have significantly more advanced
technology it is therefore not unreasonable to expect that these transmitters would
be able to detect the presence of the Earth and measure its orbit at
even greater distances. In addition to the simplicity to the receiver of
implementing this strategy it has the advantage that opposition
is typically the time when observations are easiest to make anyway.
A number of all-sky surveys that have
already been performed naturally contain tiny
``opposition surveys" within them. A full all-sky opposition survey
would require extensive time to complete with a single moderate field
of view telescope but different types of arrays might be employed instead
including some systems already under construction.

\end{abstract}

\section{Introduction}

Essentially all of the current SETI (Search for ExtraTerrestrial
Intelligence) observing programs make only brief observations
of any particular part of the sky.
This means that transmitters with low duty cycles are very unlikely
to be detected.
However, it has
been proposed by a number of authors that some type of synchronization
of transmitter and receiver can result in huge energy savings to the
transmitter by restricting transmissions to certain short times
which can nevertheless be determined by a receiver utilizing the same
synchronization scheme.
One synchronization strategy, which requires substantial effort by both the transmitter
and the receiver is to use some type of external astrophysical event
as a synchronizer. The transmitter immediately transmits when a
noteworthy event occurs
and the receiver looks for the signal after the calculated time
delay. The calculation of this delay requires the receiver
to have the ability to both
detect and localize the astrophysical event
and also precisely measure the distance to a potential transmitter.
In an earlier paper (Corbet, 1999) 
it was proposed that for this strategy gamma-ray bursts
are the best of the known potential synchronizers that are available
primarily because of their large apparent luminosities
and very brief durations. Earlier
suggestions involving for example supernovae (Tang, 1976) and novae
(McLaughlin, 1977; Makovetskii, 1980) have also been made.

In this paper
it is proposed that another timing scheme may also be
used which requires little work on the part of the receiver but
substantially more
effort by the transmitter. This scheme uses a very local
(to the receiver) astrophysical event and
is simply that potential SETI targets
should be observed when they have their maximum angular distance from
the Sun as
seen from the Earth, i.e. in the terminology often used to describe planetary
positions,
they are at ``opposition.'' Although the angles involved here may be
substantially less than 180\degrees\ (but more than 90\degrees)
the term ``opposition" is used
for convenience.
This technique requires that the transmitter must not only be able
to detect the presence of the Earth but also
be able to measure its orbit,
distance, and proper motion sufficiently accurately
for the signal to arrive at the correct time.
This scheme would
make a
beacon relatively easy
for the receiver to find while at the
same time conserving energy use by the transmitter.

\section{Why Send Synchronized Low Duty Cycle Signals?}

While an omnidirectional transmitter is the simplest to consider,
and may well be appropriate for at least some types
of ``leakage'' radiation,
this scheme is very inefficient and so is not likely to be desirable for a
beacon transmission.  If the transmitter can utilize one or more narrow
beams then large increases in efficiency can be achieved. However, if
beaming is used it will not be possible to send to all desired targets
continuously if the number of beams that can be produced simultaneously
is less than the number of targets.  Restrictions on transmission duty
cycle can also result if a system is used for more than just interstellar
transmission, for example the system might also be used for astronomical
observations.

The use of continuous transmission may also be expensive in terms of
energy consumption. One way to consider this is to compare the relative
costs of transmitting a signal to accelerating an interstellar probe
such as proposed by Bracewell (1960). This
comparison, while simplistic, is independent of the amount of energy
available in that there are energy costs associated with both these
means of exploration.
For illustration, if the Arecibo 1 MW planetary radar is
operated continuously then this corresponds to the same
energy costs as accelerating 1kg masses
to 1\% of the speed of light
approximately every 50 days.  If a beacon is operated continuously for a
long period of time then the total energy expenditure could be extremely
large (e.g. Bracewell, 1996). Hence it may well be desirable to use the
energy-saving technique of making only low duty cycle transmissions.
Further, no matter what average power budget is available to the transmitting
civilization, it may be possible to generate more intense signals
by concentrating that power into low duty cycle signals. The detectable
range of transmissions for a particular receiver sensitivity would thus
be increased.

Another factor that may lead to the use of short duty cycle
in a long running transmission program is 
the ``psychological'' factor that, the longer a program
is continued without success, the less resources a civilization
may be willing to devote to it. In this case techniques
that would both save energy costs and use less time
with transmission facilities that could be used for other purposes
would become more important.

In addition to speculation about how an extraterrestrial transmitter may be operated, other considerations suggest that searching for
low duty cycle signals should be done. These are (i) the experimental
lack of strong persistent artificial radio emitters and
(ii) the small number of deliberate terrestrial signals
that have been sent have been extremely brief.

Searches for emission at around 1.4 GHz have covered essentially the
entire sky and no persistent source has been found (see review
by Tarter, 2001). While this may simply
mean that extraterrestrial transmissions were too faint to be detected
in these surveys, or that transmissions are not being made near this
frequency, an alternative explanation may be that transmissions exist
but only have low duty cycles. Weak evidence for the existence of such low
duty cycle signals may come from the non-persistent candidate signals seen
with Big Ear (Kraus, 1979; Gray \& Marvel, 2001) and META (Horowitz \&
Sagan, 1993; Lazio et al., 2002).

The few deliberate transmissions that have been made from the Earth have
been beamed in only one direction at a time
and had extremely short transmission
durations.  A message sent from Arecibo in 1974
in the direction of the globular cluster M13
lasted only 169 s
(The Staff at the NAIC, 1974). Although somewhat longer, the ``Cosmic Call''
made in 1999 (Dutil \& Dumas, 1998) still only sent four transmissions,
each transmission of four hour duration,
to nearby stars
using the Evpatoria 70 m antenna
(Zaitsev \& Ignatov, 1999).

The clear disadvantage of low duty cycle transmissions is that initial
searches may not detect such signals.  If the transmitter is only sending
brief signals to a particular target it is thus advantageous to make
use of one or more synchronization schemes. Without synchronization it
will be very unlikely that the potential recipient will detect the signal.
Although unsynchronized signals might be detected by a receiver monitoring
the entire sky the entire time, such a system may not necessarily be employed
by the potential recipient and, even if it is, may be of reduced
sensitivity compared to a receiver with a narrow field of view.

The use of low duty cycle signaling does not necessarily
exclude the presence of an additional continuous signal. A transmitting
civilization might plausibly choose to combine both continuous low
intensity signals with brief more intense signals. The lower-intensity
signals will be accessible to receivers with sufficient sensitivity
while the low duty cycle signals will be accessible to less sensitive
receivers as long as observations are made at the correct time.  For a
transmitting civilization that presumably wishes to have its signals
detected it may well to choose a mixture of techniques including both
continuous and short duty cycle transmissions and for the short duty
cycle signals to perhaps use a variety of synchronization techniques.
This combination of different strategies would increase the overall
chance of at least one type of transmission being detected.

\section{Background to Opposition}
The concept of observing targets during opposition is related, to some
extent, to previously published work. Most of this related work either
notes the requirements on very tightly beamed transmissions or proposes
synchronization using binary systems other than the orbit of the Earth
around the Sun.

For transmission to nearby targets using short wavelengths (e.g. optical)
which results in very narrow beams it has been recognized
for some time that information
on the orbit of a target planet is required (e.g. Shklovskii \& Sagan,
1966; Kingsley, 1993).
Otherwise, for nearby targets, a narrow beam aimed at a star
may not illuminate a planet unless the beam is deliberately
defocussed.\footnote{Note that in proposing observations of anti-Sun
ecliptic longitude no particular wavelength regime is suggested and this
technique can be used at radio, optical, or other wavelengths.}

Fillipova et al. (1991) argued that even if transmitters are using radio
wavelengths they would still employ very narrow beams. These authors
proposed that the transmitter would therefore aim a continuous signal
directly at the target star. The potential receiver should thus observe
stars close to the ecliptic plane at the time of opposition when its
planet would pass through this beam. This scheme
does have some problems as it assumes that the transmitter is not
capable of detecting the presence of planets around the target star. The
resulting time spent transmitting while the target planet is not in the
path of the signal beam is thus wasteful of energy.

Pace \& Walker (1974) proposed transmission to binary stars at
observation of a particular binary phase of the recipient and, in the
opposite direction, search for transmission from a binary star system
at a particular phase. Pace (1979) extended this scheme to communication
between single stars by utilizing binary systems in angular proximity to
the line between the transmitter and receiver.  In a similar way, Singer
(1982) proposed that a transmitter could time the arrival of a signal at
the Solar System to coincide with a particular phase of the orbit
of Jupiter with respect to the transmitter.  The orbit of Jupiter was
chosen as this has the largest influence on the motion of the Sun. Singer (1980)
proposed that the four phases of maximum and minimum displacement of
the Sun/Jupiter system as viewed from the transmitting star should be utilized.

In addition to the more general considerations listed above, it is noted
that in 1924 an attempt was organized by David Todd, an associate of
Percival Lowell, to listen for artificial radio signals from Mars during
the time of opposition (Dick, 1996). This, apart from the much smaller distance
involved which makes signal travel time much less important, is very
close to the proposal to look for transmissions from around other stars
at opposition.

\section{Proposal}

It is proposed that SETI observations should be made, either of individual
targets or a general sky region, for a time period which includes
the {\em exact} moment when that target or region is at its maximum angular
distance from the Sun as observed from the Earth.  The justification
for this is: (i) the scheme provides a simple way for the receiver
to achieve synchronization - the transmitter can be expected to
want to make
it as easy as possible for its signal to be found, (ii) advances in
technology should soon demonstrate feasibility through the detection
and characterization of terrestrial mass extra-solar planets, and (iii)
for an orbit-based synchronization scheme
if the Earth is the desired target of the ETI's transmissions
(because it is the only planet in the solar system habitable zone;
Kasting, Whitmire \& Reynolds, 1993)
then the
parameters of the Earth's orbit are much more natural to use than those
of another planet or an external binary star system.

The scheme proposed here is analogous to that presented by Pace \&
Walker (1974) but is an extension to use the Sun/Earth system itself as the
binary system.  While Singer (1982) argued for synchronization using
Jupiter's orbit, as this is the major perturbation to the proper motion
of the Sun, we are now at the stage where, even with current technology,
 we will soon be able to detect nearby terrestrial planets - this
is discussed in more detail in the next section.  In principle, a binary
system has a number of phases which might be identified as important. For
example, alignment, maximum angular separation as seen by the transmitter,
and, for an eccentric orbit, apastron and periastron may also be used.
Thus, with Singer's Jovian technique and the Pace-Walker binary scheme,
for example, there is a lack of one specific orbital phase which is
definitely preferred. However, in the case of using the Earth's orbit
it appears clear that the time of opposition is most suitable. For the
receiver this orbital phase gives the minimum contamination from the
emission from the Sun. At this time the receiver and target are also, by a miniscule
amount, closest to each other which is arguably a ``psychological''
factor to justify this choice of phase.

Note that synchronizing transmissions to the Earth's orbit does
not necessarily imply that the signals can only be detected by
equipment on the Earth itself. If the beam size of the transmission
is sufficiently large then the signal may be detected, for example, by
spacecraft located elsewhere in the Solar System.  The Earth's orbit
is simply being used as a local synchronizing astrophysical event that
may be regarded as important by a civilization with ties to that planet.
It has been proposed before that beams would be made sufficiently broad,
perhaps a few astronomical units at the target, in order
to illuminate the entire habitable
region around a planet (e.g. Townes, 1993) but this suggestion is
based on an assumption that the transmitter would not know the location
of the target planet or not be able to accurately predict its location
when the signal arrives.
If the presence and orbit of a target planet are
known then, in principle, the ultimate
limiting size of the beam might be as small
as the size of the planet.  However, this is not a requirement of the
strategy proposed here and the transmitter may need to decide between
a very narrow beam, which would be as small as possible centered on the
target planet, and a broader less intense beam which could illuminate
observatories far from the Earth.

\section{Feasibility of Synchronization to Orbital Phase}

\subsection{General Considerations}

If transmissions are made once per target year then the efficiency of
the opposition technique depends on how brief a signal can be made compared to the
length of the target planet's year.  In order for the signal to arrive
at the desired orbital phase the transmitter must know precisely: (i) the
orbital period and phase of the target (e.g. the Earth), (ii) the distance
to the target, (iii) the proper motion and rate of change of the distance
to the target, and (iv) if necessary, the rate of change of the orbital
period.  We can gain some insight into the feasibility of this scheme
by considering current technology and that presently under development.
It should be kept in mind, however, that it is usually expected that
any civilization that we make contact with will have significantly more
advanced technology. This is because we ourselves only recently acquired
the ability to signal across interstellar distances and, unless the
lifetimes of communicating civilizations are very short, it is
statistically unlikely
that we would make contact with another similarly young civilization
(Shklovskii \& Sagan, 1966).  It is therefore not unreasonable to
assume that the transmitting civilization has much better technology
for detecting and measuring the parameters of terrestrial planets.
 
\subsection{Planet Finding Missions}
 
Although numerous extra-solar giant planets are now known to exist
(e.g. Marcy \& Butler, 1998; Naef et al., 2001), with the exception of the objects around the pulsar B1257+12
(Wolszczan \& Frail, 1992; Konacki, Maciejewski, \& Wolszczan, 2000)
no terrestrial mass planets have yet been found. However, this lack
of such planets may be ascribed to the insensitivity of the technique
used so far (high precision optical radial velocity measurements) and there are several missions under development which
it is anticipated will find terrestrial mass objects.
An overview of the basic techniques to be employed and some of the
missions under consideration in the search for Earth-like planets is
given by Woolf \& Angel (1998).  In NASA's program, for example, the
Space Interferometer Mission (SIM), currently proposed to launch in 2009,
is expected to have the capability of finding planets with masses not
that much greater than the Earth around nearby stars by measuring stellar
parallaxes to precisions of micro-arcseconds. Following on from SIM may
be the Terrestrial Planet Finder (TPF; Beichman, Woolf \& Lindensmith,
1999) proposed for launch in 2012. By using nulling interferometry
(Bracewell, 1978) in the infrared 
or a visible light
coronagraph
the TPF should be able to directly
observe planets around stars up to 15 pc away with sizes as small as
the Earth.  The European Space Agency (ESA) is also considering missions with
similar objectives.  ESA's GAIA mission (e.g. Gilmore et al., 1998) will
perform high precision astrometry and the Darwin mission (e.g. Penny et
al., 1998) is being investigated for later launch, perhaps in or after
2015. As proposed for the TPF, Darwin may also perform nulling interferometry
in the infrared with the primary aim of detecting Earth-like planets.
Following on from these missions, if the substantial technology demands
can be met, may be a ``Planet Imager''
mission.\footnote{See http://origins.jpl.nasa.gov/missions/missions.html}
A ``Planet Imager'' would have sufficient resolution that an
Earth-like planet could
be imaged with multiple pixels and might require arrays of TPF-like
interferometers separated by distances of thousands of kilometers.
For the special case of terrestrial size planets viewed close
to their orbital plane, the Kepler Mission 
(Koch et al., 1998; Borucki, Koch, \& Jenkins, 2001),
currently scheduled for launch in 2006, should be able to detect
these systems at much greater distances by the photometric detection
of planetary transits.

From a consideration of the missions currently being constructed
and designed it is clear that primarily
through space-based interferometry our
terrestrial planet detection abilities will soon enormously increase.
With their high precision astrometry these missions will also
provide significantly improved distance estimates.
In
addition to the expected significantly higher technology expected for
an alien transmitter, if multiple ETIs exist and are in communication with each
other then, by sharing star catalog data, they can obtain vastly
improved distance, proper motion, and velocity information. Their parallax baselines
would then be measured in parsecs rather than astronomical units and,
for the opposition technique, it is the precision with which the transmitter,
not the receiver, can measure the distance which is relevant. For
determining proper and radial motion of the target the expected larger
age of any transmitter should also enable it to be able to obtain very
precise measurements.  It is emphasized again that the discussion of
Earth's current technology level for detecting terrestrial planets is
simply to demonstrate that the proposed strategy is likely to be feasible.

\subsection{Specifics}

The minimum duration of the transmitted message ($t_m$) depends on how
well
the transmitter knows the parameters of the system transmitted
to. For example, consider a planet at a distance $D$ from
the transmitter, a mean radial velocity of $v$,
an orbital period around its star of $P_{orb}$, and a proper motion
of $\mu$. The associated measurement errors on these
parameters being given by $\delta D$, $\delta v$, $\delta P_{orb}$,
and $\delta \mu$. The signal travel time $T$ is simply $D/c$ but, for
changes in the system, the relevant timescale is $2T$, the combination
of light travel time from the receiver to the transmitter plus
the signal travel time from the transmitter to the receiver.
The first order constraints arise from the uncertainties in
the distance to the star and the orbital period and phase. The distance
uncertainty simply yields:
\begin{equation}
t_m > \delta D/c
\end{equation}
During the duration $2T$ there will be $2T/P_{orb}$ orbital cycles yielding: 
\begin{equation}
t_m > (\delta P_{orb} / P_{orb}) (2D/c)
\end{equation}
The signal duration must also be greater than the uncertainty in
the orbital phase, $\phi$. i.e.:
\begin{equation}
t_m > \delta \phi P_{orb}
\end{equation}
Second order effects arise from the radial velocity between
the transmitter and receiver and the proper motion.
The radial velocity changes the relative distance
during twice the signal travel time:
\begin{equation}
t_m > \delta v 2D/c^2
\end{equation}
The uncertainty in the change in alignment caused by the proper motion
gives
a constraint which is the time taken for the receiving planet to move
it its orbit sufficiently to bring about alignment again. For
the worst
case situation where the proper motion is entirely in the target's
orbital plane: 
\begin{equation}
t_m > (\delta \mu / 360) P_{orb} (2D/c)
\end{equation}
where $\mu$ is measured in degrees per time unit.

For illustration we may see what constraints result from the levels of
precision that are available with technology that is currently
available or is expected to soon
become available.

\paragraph{Equation 1):}
With micro-arcsecond level parallax errors, such as aimed for with
SIM and GAIA, rather small signal arrival errors could be achieved for the
nearest stars.  For example, the errors are about half an hour at 5 pc
(\sqig 15 ly) and about 3 days at 50pc (\sqig 150 ly).

\paragraph{Equations 2) \& 3):}
No extra-solar terrestrial mass planets are yet known
around normal stars and so no orbital periods have yet
been measured for such systems. Consider, however, the case of the gas giant planet
47 UMa b which has a 1089$\pm$3 day orbital period measured from 13 years
of data (Fischer et al., 2002) and is at a distance of 14.1 pc (45.9 ly).
These parameters yield an arrival time error of about 90 days. While this
is large, the size of this error, even without any improvement in the
measurement precision, decreases directly as the length of observations
increases. For example, a 1,000 year baseline alone reduces the error to
less than one day.  For 47 UMa b the current phase uncertainty given by
Fischer et al.  (2002) is about 34 days which could again be reduced by
a larger data set or more precise measurements. Even without improvement
the 90 day arrival time uncertainty is \sqig10\% of the orbital period
yielding a factor of 10 efficiency gain over a continuous transmission.
For terrestrial mass planets orbital periods may be determined in the
relatively near future, not by radial velocity measurement, but by
astrometry of the parent star (SIM/GAIA) or astrometry of the planet
itself (TPF/Darwin and perhaps the Planet Imager).

\paragraph{Equation 4):}
If a star has a radial velocity of, for example,
10 \kms\ then, at 5 pc and 50 pc
the change in the star's distance over twice the light travel time would
be
about 7 light-hours and 3 light-days respectively. If 3 \ms\ precision
radial velocity measurements are available (e.g. Butler et al., 1996) then
the distance changes can be predicted to about 10 and 100 light-seconds
respectively. The signal arrival time errors caused by velocity errors
are thus smaller than the effects of the distance errors predicted
by Equation 1).

\paragraph{Equation 5):} 
For an example of the effects of proper motion on alignment again
consider the case of 47 UMa b. The proper motion of the parent star
measured with Hipparcos is about 320$\pm$1 milli-arcseconds yr$^{-1}$.
This yields a corresponding minimum signal duration of $t_m > 1.8$ hours.
With the availability of micro-arcsecond yr$^{-1}$ precision
proper motions this would
then shrink by a factor of one thousand.

For the nearest stars even current/near future Earth technology gives
interestingly small signal arrival time errors from all parameters apart
from the effects of the uncertainty in the orbital period. However,
this figure could be reduced simply by extended observation durations.
It thus appears feasible that an extraterrestrial transmitter could use
the opposition technique for at least nearby stars. The major unknown is the maximum distance at which
this technique could be used which depends on the level of technology
available to the transmitter for finding, and measuring the orbits of,
terrestrial mass planets.

It is noted that the high precision stellar proper motion measurements
available with a SIM/GAIA type system enable very fine beams to
be utilized by a transmitter. It was earlier argued by some authors
such as Oliver (1993) that uncertainties in the proper motion
place strong constraints on the minimum size of a beam that could
be employed. However, with proper motion measured to micro-arcseconds
yr$^{-1}$
the corresponding error on the angular motion of a star at
50 pc during twice the light travel time is about 0.3 milli-arcseconds
(\sqig1.6 $\times$ 10$^{-9}$ radians). Constraints on beam width
due to stellar proper motion uncertainties are thus extremely weak
and the primary requirement for a narrow beam instead becomes a knowledge
of the orbit of the target planet.
%In fact, to be able to produce a beam this narrow would require,
%for example, an optical system with a mirror hundreds of meters
%in diameter or a radio telescope larger than the Earth.

\section{Alternative Simplified Transmission Strategy}

For an Earth-based transmitter it will be for some time
challenging to undertake the type of signaling proposed here.
Certainly the number of known extra-solar terrestrial planets will be much
smaller than the number of stars that it might be worthwhile considering
transmitting to.  Instead, a related but simpler scheme could be employed
if desired. This would be to simply transmit at the time of opposition
of a target star as seen from the Earth. In this case the detection of
the signal, if the potential receiver is not constantly monitoring for
transmissions from the Earth, relies on the receiver knowing the orbit
of the Earth.  The receiver will need to take into account the relative
proper motion of the two stellar systems during the signal travel time
as this would change the apparent time of opposition. This technique
shares some qualities of the suggestion by Shostak (1997)
to look for leakage from internal communication in binary star systems.

\section{Opposition Coverage in Current All-Sky Surveys}

Many all-sky SETI surveys such as META/BETA (Horowitz \& Sagan, 1993;
Leigh \& Horowitz, 2000) operate with a transit telescope at a fixed hour
angle. Complete coverage of the observable sky is achieved by moving the
telescope in declination at most once per day.  Thus at local midnight
(for a telescope set to an hour angle of zero) objects on the ecliptic
longitude line running through the center of the field of view will be at
``opposition."  These types of surveys have thus already automatically
performed some synchronized opposition observations.  However, the
coverage of opposition is very small compared to the total amount of
sky surveyed.

For simplicity in obtaining a very crude estimate of the amount of opposition
sky surveyed, as exact values are not important here, the offset between
the ecliptic and equatorial coordinate systems is ignored.  Consider, for
example, a detector with a ``rectangular'' field of view (FOV) $\theta$\degrees
$\times$ $\psi$\degrees\ in longitude and latitude respectively then,
at a declination of $\delta$, approximately $\theta$/(360 cos($\delta$))
of a 360\degrees\ scan would contain observations of the anti-Sun
ecliptic longitude. So for a 0.5\degrees\ $\times$ 0.5\degrees\ detector
(comparable to Project META's circular 0.5\degrees\ beam at 21 cm),
and observations made for $-30$\degrees\ $< \delta < +60$\degrees, this
FOV yields an anti-solar coverage of roughly \sqig 0.2\% or about three
days from 5 years of observations. However, this figure is just the amount of
time spent with observations
containing opposition data in the FOV and not the fraction
of data which is anti-solar.

To calculate the amount of opposition data consider that, as the FOV
crosses the anti-Sun line, this line is itself slowly moving at
slightly less than one degree per day. The time taken for the detector to
scan across the line is $\theta$/(360 cos($\delta$)) days.  The anti-solar
longitude will move about (360/365) $\theta$/(360 cos($\delta$)) degrees
during this time and the area of sky that is anti-solar during a single
scan is thus about $\theta \psi$/(360 cos($\delta$)) degrees$^2$.  For a
0.5\degrees\ $\times$ 0.5\degrees\ FOV detector at the equator this
is thus about 7$\times$10$^{-4}$ degrees$^2$ in one scan compared
to up to \sqig180 degrees$^2$ total scanned (some of the potential
observation time will typically not be usable when the telescope points too
close to the Sun). In the approximately 180 steps required to survey
90\degrees\ of declination then about 0.15 degrees$^2$ of anti-Sun sky
would be observed. So, for example, in covering the observable sky five
times there would thus be about 0.5 degrees$^2$ of exactly anti-solar
sky surveyed.  Similarly, for the planned Harvard all-sky optical survey
(Howard, Horowitz, \& Coldwell, 2000) which will have a 0.2\degrees\
(Right Ascension) $\times$ 1.6\degrees\ (declination) FOV, the
reduced size in longitude yields a
roughly 0.05\% fractional time coverage but this is compensated for
by the larger size of the FOV in latitude.

The very small size of these opposition ``micro-surveys" compared to the
total sky coverage illustrates the importance, even for an all-sky
survey, of finding the correct synchronization scheme if such is being
employed by the transmitter.  So far no all-sky survey has resulted in
a clear detection of a signal but existing data sets might be used to
investigate whether there is any excess of candidate
signals for the anti-solar observations compared to the
remainder of the dataset. For future observations it might be
be advantageous to either store the raw data from
anti-solar pointings or use a lower threshold for candidate
signals from this region and then compare the
number of candidates with other parts of the sky.  Note that the ``wow" signal reported from
survey observations made with the Ohio State University (OSU) Big
Ear telescope did not occur when the signal location was
anti-solar but more than 30 degrees from this longitude (Gray \& Marvel, 2001).

\section{Proposed Opposition Searches}

\subsection{General Considerations}

For the receiver the coordinates of the anti-Sun ecliptic longitude on
the sky are easily calculated.  For any particular time find the position
of the Sun in ecliptic coordinates (including the eccentricity of the
Earth's orbit), the line of potential target positions is then those
for which the ecliptic longitude is 180\degrees\ from the Sun and the
ecliptic latitude goes from $+$ to $-$ 90\degrees. The coordinates of
this line can then be converted to equatorial coordinates if desired. For
observing particular targets the equatorial coordinates of a target at the
estimated epoch of observations can be translated to ecliptic coordinates,
and the time when the Sun will be at that ecliptic longitude +180\degrees\
can be calculated.

\subsection{Targeted Opposition Searches}

In targeted SETI programs ``opposition'' observations may have
durations as short as desired as long as they include the exact time of
opposition. Signals sent by a competent transmitter using this technique
will be long enough to be guaranteed to include this time.  However,
for a single ground-based observatory, or satellite in low Earth orbit,
it will in general not be possible to obtain observations at exact opposition for
all desired targets in the course of
one year. It may thus be profitable to employ multiple telescopes
located at different longitudes on the Earth (as well as in both the
Northern and Southern hemispheres). If the transmission duration is
sufficiently long that observation at the exact time of opposition is not
required, then multiple observatories may not be required. However, it is
impossible to know how accurately the transmitter knows the parameters of
the Earth/Sun system which determines the minimum transmission duration
with this technique.

\subsection{All-Sky Opposition Surveys}

A full opposition survey would observe all areas of the sky
when they are anti-solar at least once.
An individual telescope might track the anti-Sun ecliptic longitude for
as long as possible at a selected ecliptic latitude, perhaps
moving to observations at a higher latitude when the original
target field falls below the horizon. The size of the
field of view of the telescope in longitude probably need not be very
large as long as the telescope is tracking the anti-solar
longitude as this dimension only gives the length of time during which a signal would
be detected. Since the anti-Sun ecliptic longitude is moving at
less than one degree per day, a 0.5\degrees\ field of view for
example can give observation
times for post-opposition longitudes of greater than half a day if the
``leading edge'' of the field of view is near the opposition
line.  
Additionally, if a signal is detected in real time
then the sky survey would be suspended for extended observations of
the region from which the signal is coming.  The coverage in
latitude, however, gives a direct reduction in the time required to complete a
full survey with all times of exact opposition covered.  This type
of survey with a limited field of view detector would be very slow as
it takes a complete year to survey a 360\degrees\ strip on the sky
at a particular latitude.
This is unlike the meridian all-sky surveys where a 360\degrees\ sky
strip is completed within a single day.
A 0.5\degrees\ field of view telescope would thus take 180 years for an
opposition survey
covering 90\degrees\ in latitude.

A survey based on the assumption that a signal would have a duration
of a particular minimum time could allow a telescope to move in ecliptic latitude
between its observation limits while remaining pointed at the
drifting opposition longitude.  This telescope might then detect signals
that had a duration longer than the time taken to scan in latitude. It
would naturally be necessary to be able to both collect data and maintain
reasonable knowledge of the telescope's pointing direction during this
``nodding" motion.

Neither scheme (nodding or staring) using a single telescope of limited
field of view is ideal for an all-sky survey.
Staring takes a long time to complete
a survey and nodding
might miss extremely brief signals. Instead it may be advantageous to
make use of the techniques considered for all-time all-sky surveys.
Such projects, whose ultimate goal is to 
continuously monitor the entire sky,
could thus have as an intermediate less ambitious
goal monitoring the anti-Sun line.
Two of these projects are the similarly named Project Argus from the
SETI League (Shuch, 1997) and the Argus Telescope being developed at
Ohio State University (Dixon, 1995).  While both projects have the ultimate
aim of continuously observing the entire sky at radio wavelengths
it will be challenging for either team to achieve this goal.

The SETI League
project aims to cover
the entire sky by using many
small diameter radio telescopes. The problem here is the
large number of dishes required, 5000 will be needed for complete
coverage but, by the middle of
2002, only 115 had been reported as operating. The OSU
telescope has a
design which employs an antenna array whose elements are combined via software
to form beams that would cover the entire sky visible from a single
site. The limitation with
this technique is the enormous computing power that is required to
cover the entire sky with both good spatial and spectral resolution
when a large number of elements, required for good sensitivity,
is employed.
By restrciting 
either of these projects to just monitoring the
anti-Sun range of ecliptic latitudes the difficulties would be much
reduced.
Either telescopes (SETI League)
or software
formed beams (OSU)
could monitor just this line to the exclusion of most of the rest of
the sky.
The number of telescopes/beams
required is determined by the beam size along the latitude line.
The small telescopes employed by the SETI League have fields of view of a
few degrees across and so the number of telescopes
required for an opposition survey
to be completed in the course of a few years does not seem unreasonable
(even though not all SETI League telescopes are steerable which
is required to track the anti-Sun line).  For an
OSU type telescope for a full survey of even one
hemisphere to be completed within a
single year would require multiple versions of this telescope at different
longitudes on the Earth. However, it is of course not a requirement
that such a survey is completed within a single year.

In principle another system that could be employed to give a very
large field of view would be if the separate components of
the Allen Telescope Array (ATA, formerly 1hT; Welch \& Dreher, 2000)
were operated and pointed
individually instead of operating as a coordinated array. However,
in contrast to the OSU and Project SETI systems, operating
the ATA in a mode suited to a rapid opposition survey results in
a substantial decrease in sensitivity. For it to be worthwhile to split
an array such as the ATA into smaller units the power transmitted
by the sender at the time of opposition compared to other
times must be sufficiently greater
than the relative reduction in sensitivity of the detector. However, this power
difference cannot be known in advance and, if transmissions are only
sent at the time of opposition then observing at this time is essential.

At optical wavelengths a telescope of the OSU
type with software formed beams cannot be made and so one or more telescopes of large field
of view are required.
One possibility may be to use a system
incorporating a Luneburg lens (also known as a
``Luneberg'' lens; Luneberg, 1944) which in principle may have
up to a 2$\pi$ steradian field of view.\footnote{A single Luneburg lens
may itself have a 4$\pi$ steradian field of view but the presence of
detectors on the lens obstructs that part of the lens.}
A system which is already continuously optically monitoring a large fraction
of the nighttime sky for astronomical purposes
is the CONCAM project (Nemiroff \& Rafert, 1999)
which utilizes CCD cameras together with fish-eye lenses.
However, the sensitivity and time resolution of CONCAM are presently
rather low and so unlikely to be useful for SETI studies.

\section{Conclusion}

With only somewhat more advanced technology than we currently posses, an
extraterrestrial transmitter could plausibly send signals timed to
arrive at a particular phase of the Earth's orbit.  The natural phase
for this is when the Earth is nearest to the apparent
position of the transmitter.  Targeted searches timed to coincide with
this phase seem feasible. All-sky searches are also possible but,
if complete exact opposition observations are to be achieved in a reasonable
amount of time, then one
or more telescopes capable of providing extensive ecliptic latitude coverage
are required.
In general, if synchronization techniques are to be used in a targeted
SETI program then it is still advocated that the external astrophysical
synchronization technique (e.g. Corbet, 1999 and references therein) should also be considered as
well as the opposition technique. Using, for example, gamma-ray bursts
as external synchronizers enables all transmissions to be made at least
as short as the brief burst timescale whereas opposition transmissions
require precise astronomical measurements of the target and the
signal arrival time accuracy decreases with increasing
distance\footnote{With an external synchronizer the effect of
larger distances is to require
longer observing times rather than transmission durations.}.
The external
astrophysical synchronizer technique is only well suited for targeted and
not all-sky searches. However, the external synchronizer technique
could be used by an extraterrestrial transmitter with approximately
our current level of technology. In contrast, the opposition technique
is most productive if a transmitter has substantially more advanced
technology. The opposition technique permits a complete all-sky or
targeted survey to be done within a predetermined period of time whereas
with the external astrophysical synchronizer, with events occurring at
random, it is not possible to predict when a particular target would
be observable.

A technique that would combine part of the philosophy of both of these
synchronization techniques would be for the transmitter to send a message
when it was at {\em its} closest point to the target. The receiver
would then calculate the expected arrival time. This strategy is,
however, only currently advocated for Earth-based transmission rather
than observing programs - for observing programs it is not even possible
yet, if we are only interested in terrestrial mass extra-solar planets,
as none have yet been detected.

\acknowledgments
I thank an anonymous referee for useful comments including the suggestion
of using Luneburg lenses at optical wavelengths.

\end{document}